\documentclass[twocolumn]{aastex631}

\newcommand{\Av}{A$_V$}
\newcommand{\jwst}{\textit{JWST} }

\begin{document}

\title{CANUCS/Technicolor: JWST Medium Band Photometry Finds Half of the Star Formation at $z>7.5$ is Obscured}

\correspondingauthor{Nicholas Martis}
\email{nicholas.martis@fmf.uni-lj}

\author[0000-0003-3243-9969]{Nicholas S. Martis}
\affiliation{University of Ljubljana, Department of Mathematics and Physics, Jadranska ulica 19, SI-1000 Ljubljana, Slovenia}

\author[0009-0000-8716-7695]{Sunna Withers}
\affiliation{Department of Physics and Astronomy, York University, 4700 Keele St. Toronto, Ontario, M3J 1P3, Canada}

\author{Giordano Felicioni}
\affiliation{University of Ljubljana, Department of Mathematics and Physics, Jadranska ulica 19, SI-1000 Ljubljana, Slovenia}

\author[[0000-0002-9330-9108]{Adam Muzzin}
\affiliation{Department of Physics and Astronomy, York University, 4700 Keele St. Toronto, Ontario, M3J 1P3, Canada}

\author[0000-0001-5984-0395]{Maru\v{s}a Brada{\v c}}
\affiliation{University of Ljubljana, Department of Mathematics and Physics, Jadranska ulica 19, SI-1000 Ljubljana, Slovenia}
\affiliation{Department of Physics and Astronomy, University of California Davis, 1 Shields Avenue, Davis, CA 95616, USA}

\author[0000-0002-4542-921X]{Roberto Abraham}
\affiliation{David A. Dunlap Department of Astronomy and Astrophysics, University of Toronto, 50 St. George Street, Toronto, Ontario, M5S 3H4, Canada}

\author[0000-0003-3983-5438]{Yoshihisa Asada}
\affiliation{Department of Astronomy and Physics and Institute for Computational Astrophysics, Saint Mary's University, 923 Robie Street, Halifax, Nova Scotia B3H 3C3, Canada}
\affiliation{Department of Astronomy, Kyoto University, Sakyo-ku, Kyoto 606-8502, Japan}


\author[0000-0001-8325-1742]{Guillaume Desprez}
\affiliation{Department of Astronomy and Physics and Institute for Computational Astrophysics, Saint Mary's University, 923 Robie Street, Halifax, Nova Scotia B3H 3C3, Canada}


\author[0000-0001-9298-3523]{Kartheik G. Iyer}
\affiliation{Columbia Astrophysics Laboratory, Columbia University, 550 West 120th Street, New York, NY 10027, USA}



\author{Gaël Noirot}
\affiliation{Department of Astronomy and Physics and Institute for Computational Astrophysics, Saint Mary's University, 923 Robie Street, Halifax, Nova Scotia B3H 3C3, Canada}


\author[0000-0001-8830-2166]{Ghassan T. E. Sarrouh}
\affiliation{Department of Physics and Astronomy, York University, 4700 Keele St. Toronto, Ontario, M3J 1P3, Canada}

\author[0000-0002-7712-7857]{Marcin Sawicki}
\affiliation{Department of Astronomy and Physics and Institute for Computational Astrophysics, Saint Mary's University, 923 Robie Street, Halifax, Nova Scotia B3H 3C3, Canada}

\author[0000-0002-6338-7295]{Victoria Strait}
\affiliation{Cosmic Dawn Center (DAWN), Denmark}
\affiliation{Niels Bohr Institute, University of Copenhagen, Jagtvej 128, DK-2200 Copenhagen N, Denmark}

\author[0000-0002-4201-7367]{Chris Willott}
\affiliation{NRC Herzberg, 5071 West Saanich Rd, Victoria, BC V9E 2E7, Canada}

\author[0009-0009-9848-3074]{Naadiyah Jagga}
\affiliation{Department of Physics and Astronomy, York University, 4700 Keele St. Toronto, Ontario, M3J 1P3, Canada}

\author{Jon Jude\v{z}}
\affiliation{University of Ljubljana, Department of Mathematics and Physics, Jadranska ulica 19, SI-1000 Ljubljana, Slovenia}

\author[0000-0001-9414-6382]{Anishya Harshan}
\affiliation{University of Ljubljana, Department of Mathematics and Physics, Jadranska ulica 19, SI-1000 Ljubljana, Slovenia}

\author[0000-0001-9002-3502]{Danilo Marchesini}
\affiliation{Department of Physics and Astronomy, Tufts University, 574 Boston Ave., Medford, MA 02155, USA}

\author[0000-0002-5694-6124]{Vladan Markov}
\affiliation{University of Ljubljana, Department of Mathematics and Physics, Jadranska ulica 19, SI-1000 Ljubljana, Slovenia}

\author[0000-0001-8115-5845]{Rosa M. M\'erida}
\affiliation{Department of Astronomy and Physics and Institute for Computational Astrophysics, Saint Mary's University, 923 Robie Street, Halifax, Nova Scotia B3H 3C3, Canada}
\affiliation{Observatorio Astronómico de Córdoba, Universidad Nacional de Córdoba, Laprida 854, X5000, Córdoba, Argentina }

\author[0009-0009-4388-898X]{Gregor Rihtar\v{s}i\v{c}}
\affiliation{University of Ljubljana, Department of Mathematics and Physics, Jadranska ulica 19, SI-1000 Ljubljana, Slovenia}

\author[0000-0002-9909-3491]{Roberta Tripodi}
\affiliation{University of Ljubljana, Department of Mathematics and Physics, Jadranska ulica 19, SI-1000 Ljubljana, Slovenia}



\begin{abstract}

We present a sample of 146 high-redshift ($z>7.5$) galaxies from the CANUCS/Technicolor surveys, showcasing photometry in every wide- and medium-band NIRCam filter in addition to ancillary HST data sampling $0.4-5 \mu m$ (22 JWST bands out of 29 bands total). Additionally, 48 (33\%) galaxies in our sample meet criteria to be classified as extreme emission line galaxies, 15 (10\%) of which are completely missed by typical dropout selections due to faint UV emission. By fitting the SEDs covering the rest-frame UV to optical at $z > 7.5$, we investigate the dust obscuration properties, giving an unbiased view of dust buildup in high-redshift galaxies free from spectroscopic follow-up selection effects. Dust attenuation correlates with stellar mass, but more strongly with star formation rate. 
We find typical galaxies at $z>7.5$ have $\sim 25 \%$ of their star formation obscured. However, since galaxies with higher star formation rates suffer more attenuation, $\sim 50 \%$ of the total star formation rate density at $7.5<z<9$ is obscured. The obscured fraction drops to $\sim 35 \%$ in our $9<z<12$ bin, possibly due to substantial dust grain growth in the interstellar medium not having time to occur. Extrapolating the decline in dust obscuration of galaxies to higher redshifts, we infer that dust obscuration should approach zero at $z > 15$, implying that epoch as when dust first forms in bright galaxies. 

\end{abstract}


\section{Introduction} \label{sec:intro}

Providing a full accounting of star formation is a critical component for building a complete understanding of galaxy evolution. Particularly, measurements of star formation conditions in the early universe place strong constraints on galaxy evolution models. One crucial component of early galaxies, which has received significant recent attention yet remains largely unconstrained, is their dust content \citep[e.g.,][]{algera23, ciesla24, cullen24, ferrara24, langeroodi_24b}. Dust plays a crucial role in properly interpreting observations of early galaxies because it preferentially attenuates light at shorter wavelengths and re-radiates it in the infrared (IR). For the highest redshift ($z \gtrsim 12$) galaxies, even with the James Webb Space Telescope's (JWST's) NIRCam \citep{rieke23}, we are limited to sampling the rest-frame ultraviolet (UV), meaning that dust may significantly affect our observations. Furthermore, by blocking the escape of UV photons, dust impacts galaxies' contribution to cosmic reionization \citep[see][for a review]{robertson21}.

Observational work has firmly demonstrated that the majority of star formation is obscured by dust up to $z \sim 4$, with the peak in the obscured fraction of $\sim 80\%$ occurring at $z=2-3$ \citep{cucciati12, rowan-robinson16, koprowski17, zavala21}. \textit{Spitzer} MIPS and \textit{Herschel} proved invaluable in performing this measurement out to $z \sim 3$, but beyond this range the poor spatial resolution, limited wavelength coverage, and poor sensitivity compared to UV-optical observations made measurements much more difficult. The Atacama Large Millimeter Array (ALMA) has enabled a revolution in this field with its vastly improved sensitivity and spatial resolution and extended the direct detection of dust emission up to and beyond $z=7$ \citep{capak15, willott15, watson15, laporte17, bakx20, bowler22}. Though a significant population of dusty star-forming galaxies exists at these redshifts \citep[e.g.,][]{laporte17, hygate23}, the total fraction of obscured star formation appears to decline toward higher redshift \citep{zavala21, algera23, traina24}, but perhaps not a quickly as previously inferred \citep{sun25}. 

Despite significant effort, detections of the dust continuum deep into the Reionization Era at $z>7$ still number only in the tens, with many coming from the forty galaxies targeted by the Reionization Era Bright Emission Line Survey (REBELS) \citep{laporte19, bakx20, bouwens22, schouws22, ferrara22, algera23, feruglio23, hashimoto23, bowler24, salvestrini24}. It is difficult to extrapolate these results to the general population due to the targeted nature of most of the ALMA observations. Most measurements come from samples preselected by their UV magnitude, potentially biasing against heavily obscured sources. Furthermore, most are very massive Lyman break galaxies, which may not be representative of the general population at these epochs. Large, blind surveys like Mapping Reionization with ALMA (MORA; \citealt{casey21}) can only achieve modest depths and cannot sample typical star-forming galaxies at these redshifts.

Recent observations with \textit{JWST} have revealed an unexpectedly large number of UV-bright galaxies in the very early universe. One proposed explanation is the absence of dust, potentially due to its expulsion during strong starbursts \citep{ferrara24}. The scenario is consistent with the non-detection of dust continuum in JADES-GS-z14-0, which places an upper limit on the dust to stellar mass ratio of $2 \times 10^{-3}$, consistent with expected yields from supernovae \citep{carniani24_alma, schouws24}. Though lower than expected given the galaxy's compact size, the dust attenuation derived from spectral energy distribution (SED) fitting is still non-negligible, with $A_V\sim0.2-0.5$ depending on the SED modeling code and assumptions \citep{carniani24_nature, helton24}. On the other hand, some models invoking bursty star formation or increased star formation efficiency suggest efficient dust production may be required to simultaneously match constraints from UV luminosity functions and properties of individual galaxies such as stellar masses and UV slopes \citep{mirocha23}.

In response to the flood of new constraints from ALMA and $JWST$, the latest generation of simulations has made significant effort to make predictions for not only the total dust mass in galaxies, but also expected number counts and observed optical depths by including varying dust models with different numerical methods \citep[e.g.][]{behrens18, lewis23, mushtaq23, choban24, esmerian24, narayanan24, zimmerman24}. There has been success in matching observed dust masses \citep{lower24}, but simultaneously matching UV and IR emission constraints has proven difficult, with evidence that sufficient resolution and the details of the stellar feedback model are important for producing realistic interstellar medium (ISM) conditions \citep{esmerian24}. Indeed, the dust properties of high-z galaxies can provide strong constraints on feedback models since they affect the amount, spatial distribution, and effective optical depth of the dust.

To truly understand the amount of obscured star-formation in the distant universe and thereby have a complete and accurate census of star formation, it is necessary to obtain an estimate of the dust content in an \textit{unbiased} sample of the intrinsically fainter \textit{normal} star forming galaxies at the highest redshifts. To address this challenge, we utilize the \jwst in Technicolor (PID: 3362) and CAnadian NIRISS Unbiased Cluster Survey (CANUCS; \citealt{willott22})  which together provide deep imaging with every \jwst NIRCam medium- and wide-band filter in addition to ancillary \textit{Hubble Space Telescope} (\textit{HST}) data, achieving space-based photometry in an impressive 29 bands. This effectively provides a low resolution spectrum for a large sample of high-z galaxies, with $R \sim 15$ covering $0.4-5 \mu$m. 
The wide wavelength range and higher spectral resolution of the medium bands importantly can differentiate strong emission lines from continuum and with the large wavelength coverage facilitate much more accurate estimates of dust extinction. The exquisite photometry allows us to account for dust via SED modeling and thus estimate the amount of obscured star formation without an expensive direct dust continuum detection or potentially biased selection for spectroscopy, albeit with the requirement that the assumed form of the dust law that is fit is indeed representative of its form at high-z.

This paper is organized as follows. In Section \ref{sec:data} we present the data utilized in this work. Section \ref{sec:analysis} describes the sample selection and analysis. Section \ref{sec:results} showcases our results. In Section \ref{sec:Discussion} we compare with other results and finally draw conclusions in Section \ref{sec:Conclusion}. Throughout the paper we assume a standard $\Lambda$CDM cosmology with $\Omega_\Lambda = 0.7$, $\Omega_M = 0.3$, and $H_0 = 70$ km s$^{-1} Mpc^{-1}$ as well as a Chabrier initial mass function \citep{chabrier03}. All magnitudes are in the AB system.  

\section{Data} \label{sec:data}

We draw our sample from the JWST in Technicolor survey (Program ID 3362), which, along with CANUCS \citep[Program ID 1208][]{willott22}, covers the flanking fields of three Hubble Frontier Fields clusters (Abell 370, MACS J0416.1-2403, and MACS J1149.5+2223) with comprehensive NIRCam imaging. These fields possess homogeneous imaging in \textit{every} wide- and medium-band NIRCam filter along with two narrow-band filters (F164N and F187N) and ancillary {\tt HST} ACS imaging from HFF program \citep{lotz17}. The $3\sigma$ limiting flux densities in a $0".3$ aperture range from $\sim 2-4$ nJy in wide filters to $\sim 4-8$ nJy in medium filters. Complete details of depth and completeness will be provided by Sarrouh et al. (in prep.). The CANUCS/Technicolor image reduction and photometry procedure is fully described in \citet{noirot23, willott24, asada24}, whereas details on PSF measurement and homogenization can be found in \citet{sarrouh24}. Briefly, we begin the data reduction process with a modified version of the Detector1Pipeline (calwebb\textunderscore detector1) stage of the official STScI pipeline. Astrometric alignment of the different exposures of JWST/NIRCam to HST/ACS images, sky subtraction, and drizzling to a common pixel scale of 0.04$''$ utilize the grism redshift and line analysis software for space-based slit-less spectroscopy (\texttt{Grizli}; \citealt{brammer21}). PSFs are extracted empirically by median stacking bright, isolated, non-saturated stars. Object detection and segmentation are performed on the $\chi_{mean}$ detection image created using all available NIRCam and {\tt HST} ACS images. Photometry is performed on images convolved to the F444W resolution with the \texttt{Photutils} package \citep{bradley22}. Data for CANUCS may be found at \href{https://doi.org/10.17909/ph4n-6n76}{doi:10.17909/ph4n-6n76}.

\begin{figure}[ht]
\plotone{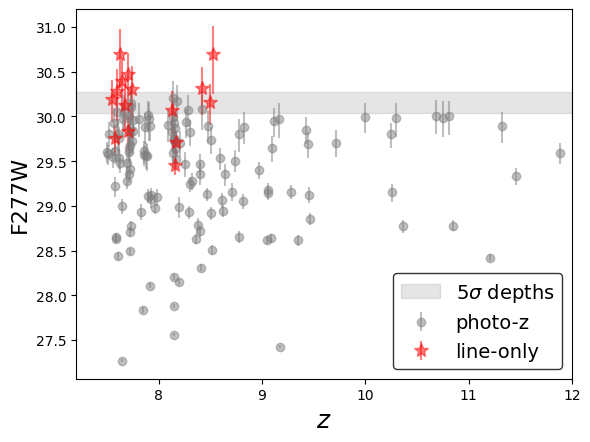}
\plotone{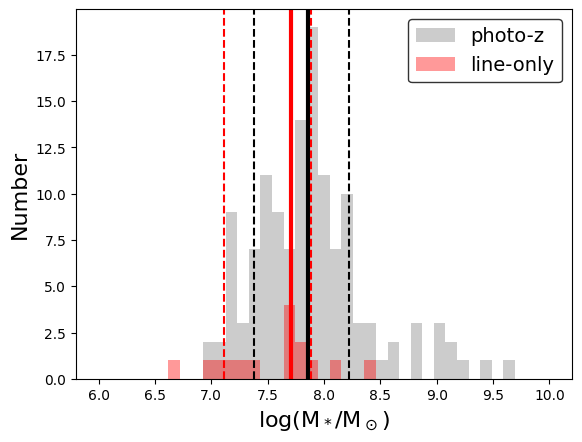}
\caption{Top: F277W magnitude versus redshift for the photo-z selected sample (gray) and additional ELG sample that is missed by the photo-z selection (red, see Section 3.2). Bottom: Stellar mass distributions for the photo-z selected (gray), and line-only (red) samples. Solid vertical lines indicate the distribution medians, whereas dashed lines show the 16$^{th}$ and 84$^{th}$ percentiles.
\label{fig:mag-z}}
\end{figure}

 \section{Sample Selection} \label{sec:analysis}
\subsection{Photo-z Selected Galaxies at $z>7.5$}

 As our fiducial high-redshift sample, we apply a similar selection to that used by \citet{willott24} to identify $z>7.5$ galaxies with high fidelity for the high-z CANUCS luminosity function, albeit opting for a more complete than pure selection by allowing slightly lower signal-to-noise ratios in F277W. We require a dropout in the F090W filter and a confident high-z photometric solution from EAzY \citep{brammer08}. The full selection is as follows:

 \begin{equation}
     S/N(F277W) \geq 5
 \end{equation}

 \begin{equation}
     S/N(F090W) < 2
 \end{equation}
 
 \begin{equation}
     z_{phot,ML} \geq 7.5
 \end{equation}

 \begin{equation}
     z_{phot,16} \geq 6
 \end{equation}

Here $z_{phot,ML}$ refers to the maximum likelihood redshift from \texttt{EAzY} and $z_{phot,16}$ is the $16^{th}$ percentile of the redshift probability distribution. When running EAzY we use modified templates based on the \texttt{binc100z001age6\_cloudy\_LyaReduced} template from \citet{larson23} that include stronger [OIII] emission lines to better match recent observations and the empirically-calibrated intergalacic medium/ circumgalactic medium attenuation curve of \citet{asada25} which has been shown to reduce redshift bias for Epoch of Reionization galaxies. By requiring a non-detection in F090W and high confidence in a photometric redshift solution above $z=6$, we exclude low-redshift interlopers from our sample. This selection results in 251 galaxies, which is reduced to 131 secure sources after visual inspection. Most of the removed sources are spurious detections or objects with insufficient signal to perform reliable SED fitting. We refer to this as the photo-z selected sample throughout the paper. 

\subsection{Emission Line Selected Galaxies at $z>7.5$}

In addition to this more typical high-z selection, we also incorporate color-selected emission line galaxies (ELGs) identified by flux excesses in the photometry caused by the [OIII]+H$\beta$ complex. The ELG selection ensures a robust photometric redshift solution and potentially allows us to recover galaxies whose rest-frame UV emission is too faint for a secure photo-z selection using traditional methods \citep{cava15, lumbreras-calle19}. The process and specific color combinations for the selection are described in detail in \citet{withers23}. Briefly, template spectra from the \texttt{Yggdrasil} stellar population synthesis code \citep{zackrisson11} were used to derive color selections to identify emission line excesses in synthetic photometry. Completeness and contamination tests led to the inclusion of a S/N$\geq 4$ cut in line emission and an average S/N $\geq2$ cut on the underlying continuum bands. Follow-up spectroscopy of 15 objects presented in \citet{withers23} confirmed excellent redshift precision and accuracy from this technique, with $|z_{phot}-z_{spec}|/(1+z)=0.0087$ and no outliers, albeit in a sample at lower redshift which was further constrained by the presence of H$\alpha$ in the observed photometry. Additional details will be given in Withers et al. (in prep.).

Our selection identifies 67 objects, which is reduced to 48 after removing potentially spurious sources and sources which are too faint to retrieve constrained properties from SED fitting via visual inspection.  We find 15/48 ($31\%$) galaxies in the ELG sample are missed by the photo-z selection, bringing the total number of unique sources to 146. These missed ELGs generally have weak rest-UV emission, and are thus excluded by the F277W S/N cut. Two are excluded for having photo-z posteriors from \texttt{EAzY} which extend to lower redshifts. Throughout the paper we will distinguish between ELGs which are included in the photo-z sample as the continuum+line sample (owing to the F277W UV continuum and Lyman break detections, 33 galaxies), and "missed" ELGs as the line-only sample (15 galaxies).

Figure \ref{fig:mag-z} shows the F277W magnitude versus photometric redshift for our continuum high-z selection (gray) as well as the additional 15 sources identified by our ELG selection (red). Indeed, the line-only subsample probes fainter UV magnitudes than the photo-z selection, enabling the detection and characterization of fainter galaxies. Note these "missed" ELGs are important as the lack of UV continuum can be caused by strong dust attenuation in some sources, not only because they are always intrinsically fainter.  We will show some of the dustiest galaxies are among the emission line only sample. The bottom panel of Figure \ref{fig:mag-z} shows that the fainter UV magnitudes do not necessarily correspond to lower stellar masses. The stellar mass distributions largely overlap, with the line-only sample missing the tail to higher masses seen in the photo-z sample. We further explore the relationship between these samples below.

\begin{figure*}[h]
\plotone{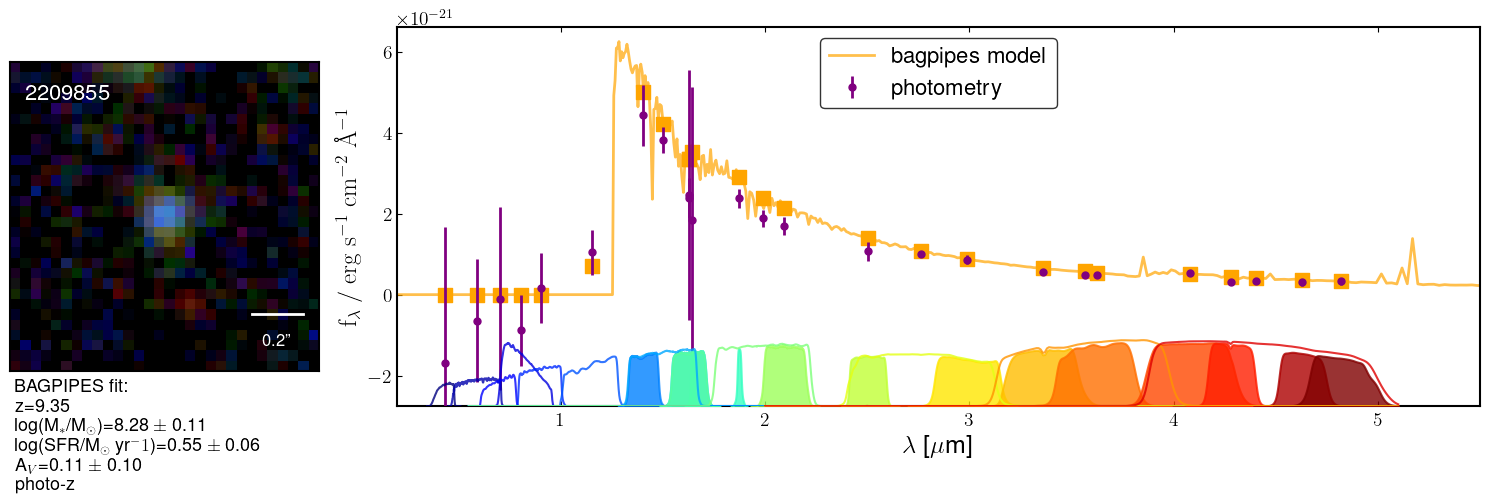}
\plotone{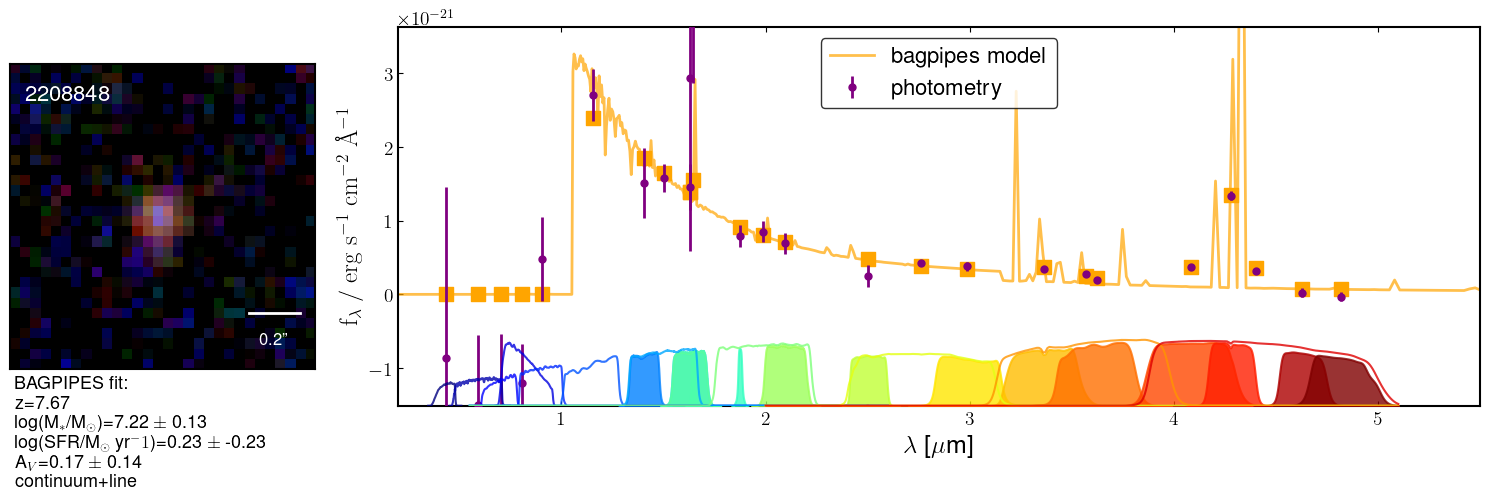}
\plotone{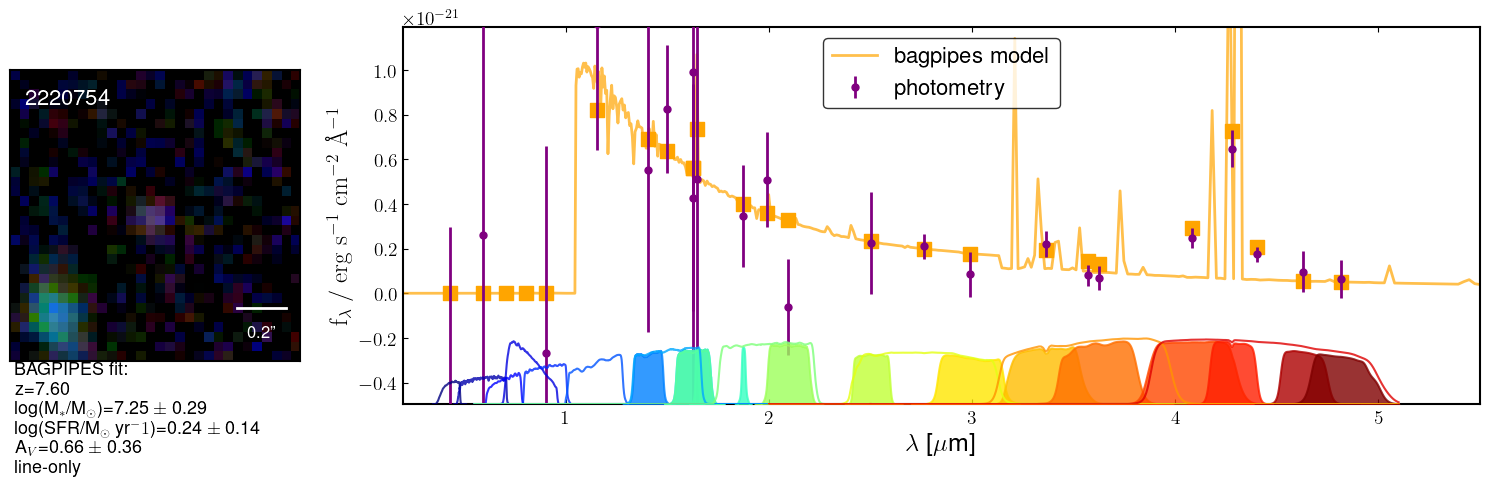}
\caption{RGB (F444W, F277W, F200W) image stamps, SEDs (purple), and \texttt{BAGPIPES} models (orange) with accompanying physical properties for three example galaxies. Filter curves corresponding to the photometry are shown on the bottom, with wide-band filters outlined, and medium-band filters filled. There is one example each from the photo-z sample, continuum+line sample, and line-only sample as labeled.
\label{fig:SEDs}}
\end{figure*}

\subsection{SED Modeling}
We perform SED modeling of the HST+JWST photometry with the \texttt{BAGPIPES} code \citep{carnall18} to derive physical properties for our sample. We assume a double power law star formation history, Calzetti \citep{calzetti00} dust attenuation law, and Chabrier \citep{chabrier03} initial mass function. \citet{markov24} recently used JWST prism spectra to measure the attenuation curves for galaxies up to $z=12$. They find that for $z>6$ galaxies, the empirically measured attenuation curve is similar to the Calzetti curve, with a shallow slope and no UV bump. We thus adopt the Calzetti curve to facilitate comparison with other works and show how the measured attenuation values differ when assuming different dust laws, priors, and SED fitting methodology (\texttt{DENSEBASIS}, \citealt{iyer19}) in the appendix. The redshift is set to the photometric redshift from \texttt{EAzY} with a Gaussian prior of width 0.05. The exquisite Technicolor medium-band photometry strongly constrains the shape of the continuum while overlapping medium- and wide-bands quantify the contribution of emission lines. Figure \ref{fig:SEDs} shows three example SEDs, one each for our photo-z, continuum+line, and line-only subsamples. Photometry is shown in purple and the corresponding \texttt{BAGPIPES} fit in orange. We also show RGB cutouts and the measured stellar population properties for each object.  

\begin{figure*}[h!]
\includegraphics[width=\textwidth]{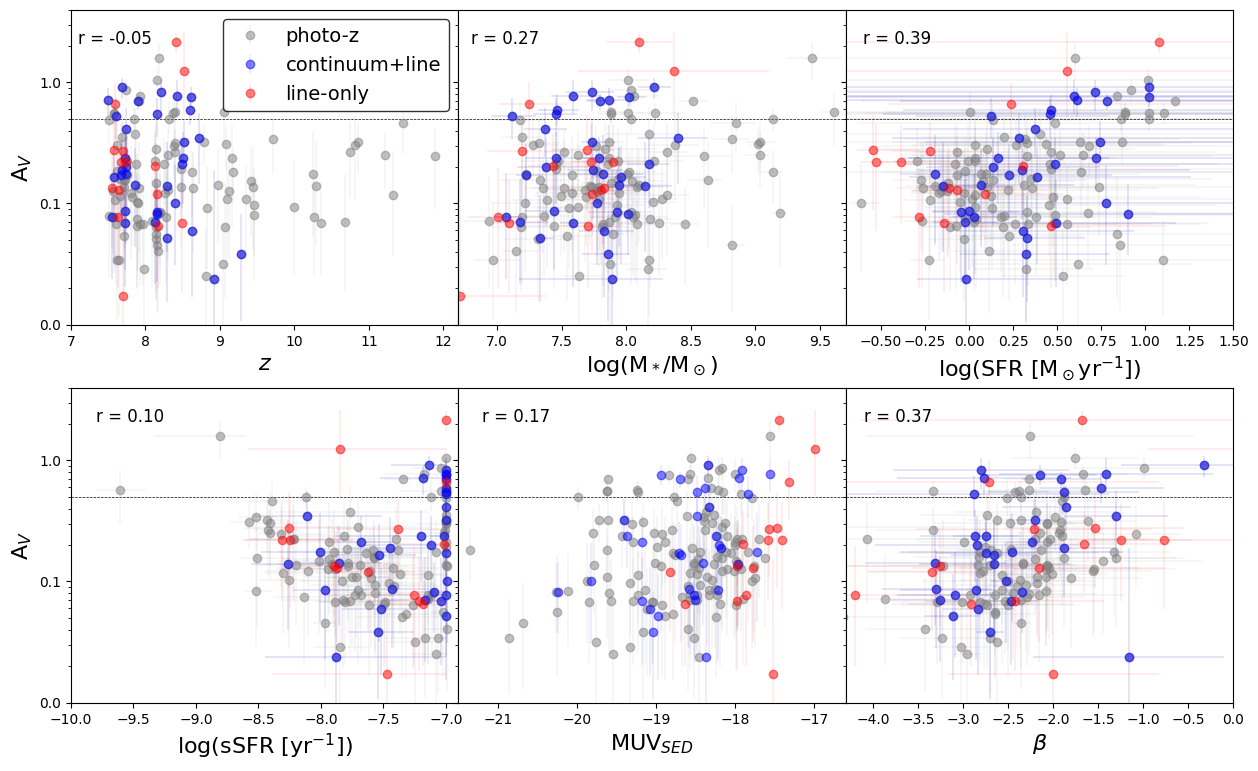}
\caption{Correlation of total V-band attenuation {\Av} with physical quantities. The photo-z sample is shown in gray, continuum+line sample in blue, and line-only sample in red. The Pearson r correlation coefficient between the two quantities is shown in the top left of each panel. The horizontal dashed line indicates $A_V=0.5$.}
\label{fig:Av}
\end{figure*}

\begin{figure*}[h]
\includegraphics[width=\textwidth]{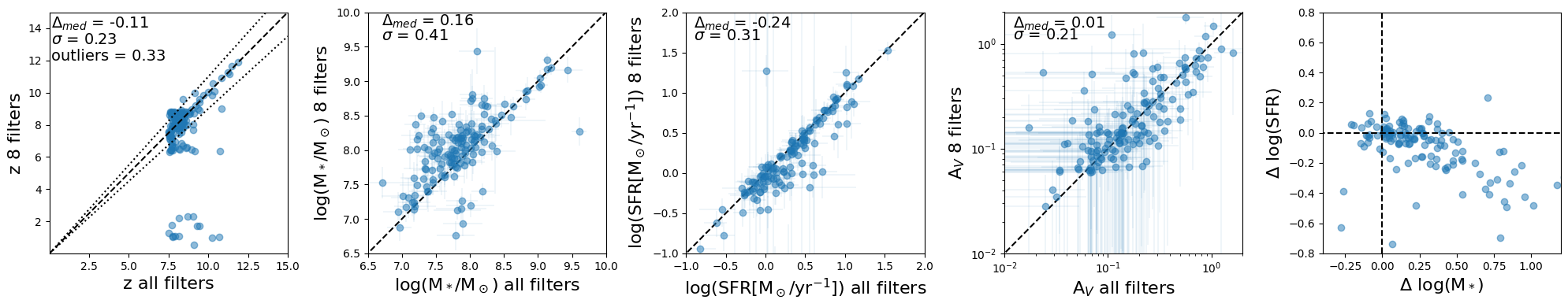}
\caption{Comparison of SED fitting results using the full Technicolor filter set versus a typical eight filter set. The median offsets ($\Delta_{med}$) and scatter ($\sigma$) between the two fits are printed in each panel. The redshift comparison plot also displays the outlier fraction, defined as $\Delta z/(1+z)>0.1$. Equal values are indicated by dashed lines. The dotted lines in the first panel indicated the defined outlier range. The rightmost panel shows the correlation between the offsets in SFR and stellar mass.}
\label{fig:broadband}
\end{figure*}

The analysis of $\beta$ slopes for the full CANUCS sample will be presented in Felicioni et al. (in prep.). Here, we fit a power law to the observed photometry utilizing all bands whose pivot wavelengths lie between rest-frame $1215-3200$ \r{A}. At $z=8$, this corresponds to 10 \jwst medium and wide filters (F115W, F140M, F150W, F162M, F182M, F200W, F210M, F250M, F277W; the two narrow bands F167N and F187N are included in the fit, but are shallower and provide less constraint). 

\section{Results} \label{sec:results}

\subsection{How Many Galaxies Exhibit Significant Dust Obscuration at $z > 7.5$?}

We first investigate how dust obscuration in our sample relates to other galaxy physical properties. Figure \ref{fig:Av} shows the dependence of {\Av} on redshift, stellar mass, star formation rate (SFR), specific star formation rate (sSFR), absolute UV magnitude (MUV), and $\beta$ for our photo-z sample (gray), continuum+line sample (blue), and line-only sample (red). Stellar mass, SFR, and {\Av} are the outputs from the \texttt{BAGPIPES} fit, MUV is measured from the best fit \texttt{EAzY} template, and $\beta$ is calculated as described above. While many high-z sources are consistent with little to no dust obscuration, we find a significant population of objects with {\Av} $> 0.5$. Specifically, 15\% of the continuum sample, and 25\% of the continuum+line sample, and 20\% of the line-only sample meet this criterion. The median statistical uncertainties on the derived {\Av} are 0.12 mag for the continuum sample, and 0.15 mag for the line only sample, though systematic uncertainties are significantly higher (see Appendix). 

As reported in \citet{willott24}, the CANUCS fields do not exhibit a high density of UV-luminous galaxies at high-z, thus our sample does not include a large number of objects with extreme stellar masses derived in some other recent works \citep[e.g.,][]{labbe23, finkelstein24, xiao24}. We find eight galaxies with log(M$_*$/M$_\odot)>9$. This is most likely due to cosmic variance since \citet{willott24} also do not find massive galaxy candidates in the CANUCS cluster pointings. Another possibility, though likely subdominant for this sample, is that the more accurate photometric redshifts enabled by the medium bands better exclude low-redshift interlopers that may be interpreted as more massive galaxies at higher redshift in surveys with fewer filters, as has been found at lower redshift \citep{sarrouh24}. However, the medium-band coverage excluding SED solutions with high stellar masses that would otherwise be unconstrained with only broadband photometry \citep{desprez24} could also contribute (see next section).

Similarly, SFRs of our sample are modest, falling below 15 M$_\odot$ yr$^{-1}$. The artifact of galaxies being limited to $10^{-7}$ yr$^{-1}$ in sSFR (visible in the bottom left panel of Figure \ref{fig:Av}) comes from the adopted choice of time resolution of 10 Myr for the SED modeling. The inclusion of the line-only subsample extends the range of MUV reached by this work to -17 mag demonstrating the efficacy of the color selections to identify intrinsically UV-faint galaxies. 

In each panel of Figure \ref{fig:Av} we report the Pearson correlation coefficient of {\Av} with the indicated quantity. We do not observe strong correlations between \Av\  and MUV, but we identify mild trends toward increasing \Av\ with increasing stellar mass and SFR, with correlation coefficients of 0.26 and 0.39. The trends are in line with findings in both the local Universe from the Sloan Digital Sky Survey \citep{brinchmann04, gilbank10} and cosmic noon \citep[e.g.,][]{sawicki12, martis16, whitaker17}. We also recover the expected correlation between $\beta$ and dust obscuration (r=0.37), but with substantial scatter. 

\subsubsection{Dust Properties from Medium versus Broadband Photometry}

To test the hypothesis that medium-band photometry affects how dust properties are derived, we repeated our SED fitting procedure including only the F090W, F115W, F150W, F200W, F277W, F356W, F410M, and F444W bands (the set used in the CANUCS cluster fields and representative of other extragalactic surveys). Figure \ref{fig:broadband} compares the photometric redshifts, stellar masses, SFRs and {\Av} derived using the two filter sets. Photometric redshifts change significantly, with a third of the sample having significantly different solutions, defined by $\Delta z/(1+z)>0.1$. Moreover, 14 (10\%) galaxies are moved from $z>7.5$ to $z \sim 1-2$ without the medium-band filters, showing that the extra filters significantly improve completeness of photometric samples at high-z. This is particularly true for faint galaxies, in agreement with recent findings from \citet{adams25}. We find that without the medium bands, stellar masses are systematically overestimated by $\sim 0.15$ dex and SFRs are underestimated by $\sim 0.25$ dex.  Moreover, we show in the rightmost panel, that the biases in stellar mass and SFR are anti-correlated, meaning that without the medium bands sampling the [OIII]+H$\beta$ complex, SED modeling prefers solutions with elevated continuum to account for the line-flux, resulting in overestimated stellar masses. Stellar mass functions at high-z may therefore be overestimated if measured from only broadband photometry. The behavior of {\Av} is more complicated. At low attenuation levels, the broadband-only fitting slightly overestimates {\Av}, whereas at higher values it underestimates {\Av}, though in both cases usually within the errors. Coincidentally, this leads to no significant median offset in the sample as a whole.

\subsection{The Fraction of Obscured Star Formation}
Beyond understanding the fraction of early galaxies that exhibit significant dust obscuration, we wish to determine the fraction of star formation that is obscured within individual galaxies. To estimate the fraction of obscured star formation, we take the following approach. First, we use the stellar population properties from the \texttt{BAGPIPES} SED fit to produce an intrinsic, unobscured model spectrum. We calculate the NUV-based SFR using the calibration of \citet{hao11, murphy11}, given below
\begin{equation}
    \mathrm{log} SFR[M_\odot yr^{-1}] = \mathrm{log}(L_{NUV}[erg s^{-1}])-43.17
\end{equation}
from both the intrinsic and obscured model spectra (assuming the Calzetti attenuation curve) and use the ratio of UV to total SFR to calculate the obscured fraction, $f_{obs}$, of the SFR in each galaxy. We verified that the dust-corrected NUV-based SFRs derived in this manner are broadly consistent with those derived from SED modeling. 

\begin{figure*}[h!]
\includegraphics[width=\textwidth]{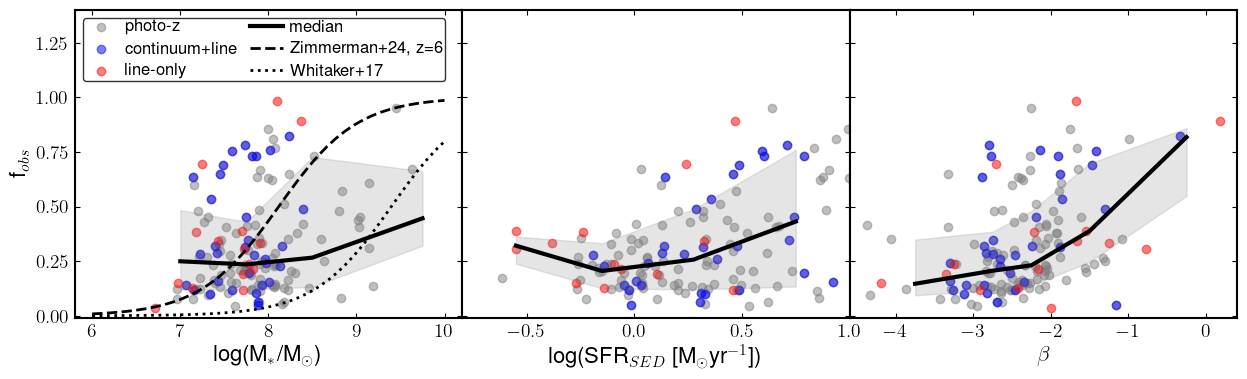}
\caption{Fraction of obscured star formation, $f_{obs}$, versus stellar mass, SFR, and $\beta$ with the continuum, continuum+line, and line-only samples indicated in gray, blue, and red, respectively, The running median is shown as a black curve with $1-\sigma$ dispersion as a shaded gray region. Relations between $f_{obs}$ and stellar mass from the literature are shown as doshed and dotted lines for comparison.}
\label{fig:fobs3}
\end{figure*}

Figure \ref{fig:fobs3} shows $f_{obs}$ as a function of SFR, stellar mass, and $\beta$. We observe tentative trends of increasing $f_{obs}$ with all three quantities, but with significant scatter. Our observed relation between $f_{obs}$ and stellar mass is shallower than that derived at $0.5<z<2.5$ \citep{whitaker17}, while also exhibiting higher obscuration values at low stellar masses. We find a closer match to recent results from the SIMBA cosmological simulations at $z=6$, though again a more shallow relation than predicted. The photo-z, continuum+line, and line-only samples occupy similar distributions of $f_{obs}$, though most of the line-only sources exhibit fairly low $f_{obs}$ along with the relatively low SFRs in these systems. Significantly, we observe 28 sources (19\% of the full sample) with $f_{obs}>0.5$. That is, the majority of star formation in roughly one in five galaxies at $z>7.5$ is strongly obscured by dust. As shown by the completeness tests in \citet{willott24}, our detection fraction remains high at faint UV magnitudes because detection is performed on a $\chi_{mean}$ image with so many filters. Even without the extra medium-band filters provided by Technicolor and more stringent selection, they reach 80\% completeness at $M_{\\UV}=-19$ and $z \sim 8$ in these fields. Their sample in these fields contains a total of 77 galaxies compared to our 146, meaning that in addition to the more lenient S/N cut employed here (our original selection of 251 galaxies is reduced to 120 using exactly the \citealt{williams24} criteria), the long-wavelength medium-band filters and addition of our line-only sample significantly improve the completeness of high-redshift galaxy samples. These measurements therefore constrain the dust content of a truly representative sample of galaxies in the early universe. 

\begin{figure}[h]
\plotone{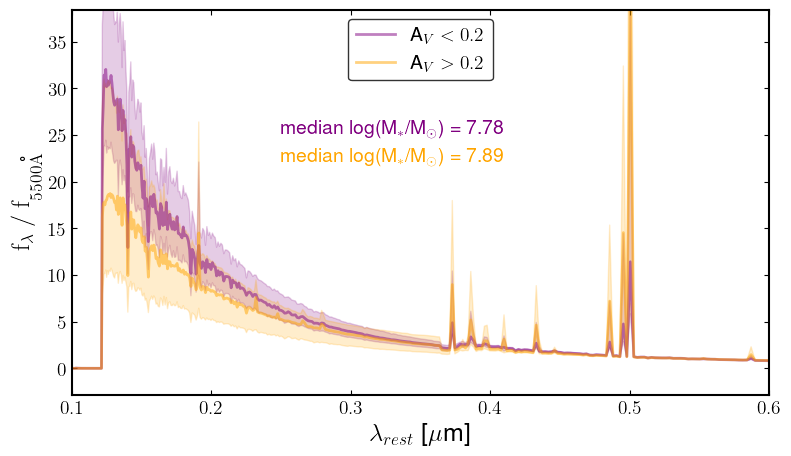}
\plotone{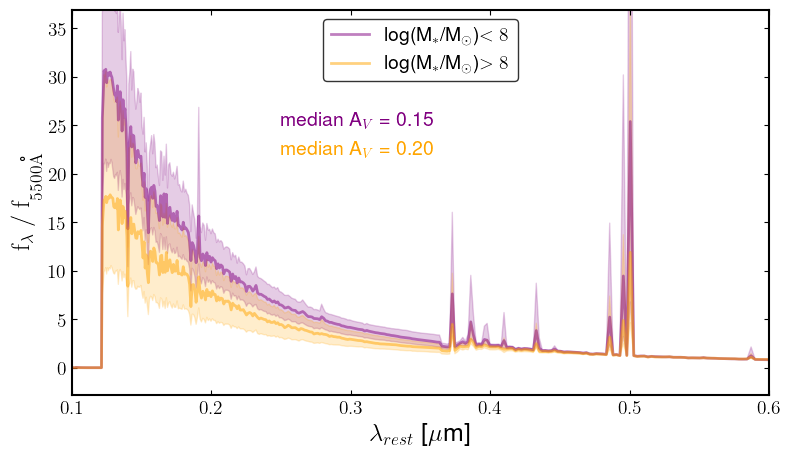}
\caption{Median model spectra from \texttt{BAGPIPES} when splitting our sample into two bins normalized to rest-frame $5500 \AA$. The top panel shows the median models for low (blue) and high (orange) dust content. The bottom panel shows the median models for low (blue) and high (orange) stellar mass.}
\label{fig:models}
\end{figure}

To illustrate the significant difference in SED shape for the more heavily dust-obscured galaxies in our sample, Figure \ref{fig:models} shows the median model spectra from \texttt{BAGPIPES} when splitting our sample into two bins normalized to rest-frame $5500 \AA$. In the top panel, our sample is divided by {\Av}, and in the bottom by stellar mass. The low dust and low stellar mass sub-samples show steeper UV-$\beta$ slopes. The low-mass and higher dust sub-samples show stronger emission lines relative to the normalized continuum.

\section{Discussion} \label{sec:Discussion}
\subsection{Evolution of Total Obscured SFR at $z >7.5$}
We have shown that $z>7.5$ galaxies exhibit a wide variety in their dust obscuration properties and how this translates into observable quantities frequently used to characterize high-z galaxy populations such as MUV and $\beta$. We now consider how these population properties translate to the global SFR density measurements by comparing the summed SFR$_{UV}$ and SFR$_{SED}$ across each sample. After accounting for dust attenuation from our SED modeling, we find $\sim40\%$  of star formation in the continuum sample is obscured, whereas this reaches $\sim50\%$ for the continuum+line sample, and $\sim60\%$ in the 15 "missed" line-only sources. 

\begin{figure*}[h]
\plottwo{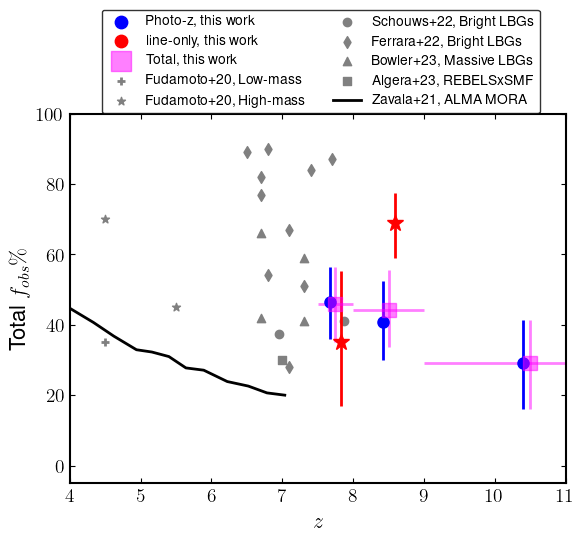}{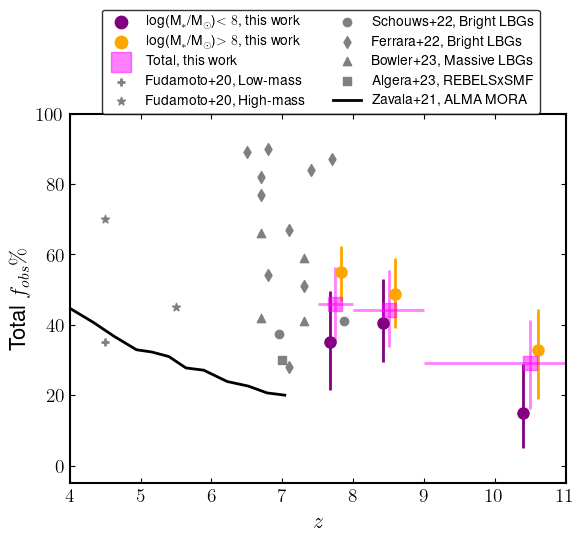}
\caption{Fraction of star formation that is obscured as a function of redshift. The left panel splits the sample into the continuum photo-z selected sample and line-only samples. The right panel splits the sample into bins of low and high stellar mass at $log(M_{*}/M_\odot) = 8$. Bin widths are indicated on the magenta points showing the total for each bin. Error bars are obtained by recalculating $f_{obs}$ adopting the $16^{th}$ and $84^{th}$ percentiles of {\Av} for each galaxy.}
\label{fig:fobs-z}
\end{figure*}

Figure \ref{fig:fobs-z} shows how $f_{obs}$ for our photo-z, line-only, and full samples evolves with redshift in comparison with other results from the literature. Errors are obtained by recalculating $f_{obs}$ using the $16^{th}$ and $84^{th}$ percentiles of the {\Av} posteriors for each galaxy. At $z=7-9$, we observe $f_{obs}$ in line with most previous estimates in the $z \sim 7$ range from the ALMA REBELS and MORA programs \citep{zavala21, schouws22, ferrara22, bowler24, algera23} and slightly lower redshift from \citet{fudamoto20}. This is perhaps an unexpected result. The ALMA sources with dust continuum detections are largely drawn from follow-up observations of the brightest Lyman break galaxies, which are potentially biased to the most massive or heavily star forming systems. Indeed, the REBELS sample has stellar masses $\log(M_{*}/M_\odot) \gtrsim 9.5$, whereas the median stellar mass of our sample is $\log(M_{*}/M_\odot) = 7.6 \pm0.2$. We thus show for the first time that half of all star formation is obscured up to $z=9$, but that this is driven by the more massive and heavily star-forming galaxies in our sample (which is still overall low-mass) being preferentially more obscured. This is shown more clearly by looking at $f_{obs}$ versus redshift after subdividing our sample into low- and high-mass with the boundary at $\log(M_{*}/M_\odot) = 8$ (right panel of Figure \ref{fig:fobs-z}). The low-mass subsample has a median $f_{obs} = 0.23$, while the high-mass subsample has median $f_{obs} = 0.32$. 

Alongside the general consensus that most low mass galaxies in the EoR exhibit low levels attenuation, several recent studies point to the existence of a population with non-negligible attenuation. \citet{sandles24} use spectroscopy from the JWST Advanced Deep Extragalactic Survey (JADES) ton investigate the Balmer decrement in low-mass galaxies from $4<z<7$. While most of the sample is consistent with little to no nebular dust attenuation, they do find trends of increasing attenuation with increasing stellar mass and SFR, in line with our results. Detailed studies of individual objects at $z > 8$ have uncovered a handful of spectroscopically confirmed  objects with $A_V \sim 1$ \citep{markov23, harshan24, ma24} accompanied by significant star formation. On the more extreme end, there have been several claims of massive ($M_* \sim 10^{10}M_\odot$), heavily obscured galaxies at $z \sim 8$ \citep{akins23, rodighiero23}. So while perhaps low in number, the most active sites of star formation even in the very early universe may be heavily obscured, leading to significant obscuration of the total star formation rate density.

At lower redshift (mostly $z<2$), \citet{bisigello23} report the finding of a population of dusty dwarf galaxies (median stellar mass $10^{7.3} M_\odot$). They exhibit extraordinary levels of dust attenuation given their low mass, with median {\Av}=4.9. These sources are too red for $HST$ and too faint for existing FIR wide-field surveys and so have been missed before \jwst. Their comparative rarity to bluer star-forming dwarfs may point to the possibility of a short evolutionary stage in which dust enshrouds very recent star-formation before being destroyed or expelled from the shallow potential well. One heavily obscured dwarf galaxy was recently spectroscopically confirmed at $z=4.88$ with stellar mass $\sim 10^{8} M_\odot$, {\Av}=2.2-3.3, and significant metal enrichment \citep{bisigello24}. Thanks to the aid of gravitational lensing, a similar $2\times10^{9} M_\odot$ stellar mass galaxy at $z=4.27$, MACS0717\_Az9, has a detected dust continuum and directly measured fraction of obscured star formation of $\sim80\%$ \citep{pope17, pope23}. Detecting similar sources at higher redshift, if they exist, will prove extremely difficult, so further investigating this population of low-mass, dusty galaxies may be the best way to study how dust might accumulate in early galaxies.

Recent simulations of galaxy evolution seem to point to detailed modeling of ISM processes as being key to reproducing the properties of epoch of reionization galaxies. Using an updated version of the SIMBA simulation which includes prescriptions for dust and molecular hydrogen directly in the simulation rather than via post-processing, \citet{jones24} find the inclusion boosts early star formation and dust to gas ratios. Spatially resolving the ISM also appears to play an important role. Two recent studies with similar prescriptions for dust and predicted dust masses for reionization era galaxies \citep{lewis23, esmerian24} predict different levels of UV extinction, with lower resolution corresponding to lower attenuation. In this case it is likely that by spreading the dust over a larger surface area, the effective optical depth is reduced. Similarly, investigations using the high resolution FIRE \citep{ma18, ma19} and FirstLight \citep{mushtaq23} find lower dust attenuation for a given dust mass than other models. By resolving the effect of stellar feedback on the ISM, these simulations can produce variations in optical depth across a galaxy, including channels through which UV radiation can escape, lowering the total effective optical depth. Measurements of attenuation in high-z galaxies may therefore be one of the best ways to distinguish between models of stellar feedback in the extreme environments of high-z galaxies.

\subsection{Timescale of Dust Production}

The $f_{obs}$ drops in in our highest redshift bin, albeit within the error of lower redshift bins, perhaps pointing to a limit in the speed in which dust can effectively form in typical galaxies at this epoch. Fitting a linear relation to our observed obscuration fraction versus redshift trend predicts the obscuration fraction to reach zero at $z \sim 15$. Such an extrapolation obviously suffers from incompleteness in our highest redshift bin, and does not reflect potentially rapidly changing conditions in the first few hundred Myr of cosmic history, but is informative in the sense that we can now point to the earliest galaxies in which we predict dust to be observable. Through an analysis of $\beta$ slopes of $z>8$ galaxies, \citet{cullen24} also find that moderate attenuation observed at $z<10$ is mostly absent at higher redshifts, with $\beta$ consistent with dust-free stellar populations. Similarly, \citet{toyouchi25} find with self-consistent, resolved modeling of stars, gas, metals, and dust that the $z\sim12$ luminosity functions can be reproduced by their model only if the predicted significant effect of dust attenuation is ignored. The low dust obscuration at $z>10$ is also consistent with the model of \citet{ferrara24}, who argue that radiation-driven outflows in very early galaxies should be able to clear dust, increasing apparent UV brightness. The combined evidence points to a rapid change in galaxies' dust properties in the first few hundred million years of cosmic history.

\citet{langeroodi24a, langeroodi_24b, ciesla24} find a correlation between {\Av} and inverse sSFR, which provides a way to determine the time delay between the onset of star formation and significant dust production since 1/sSFR is a close proxy for stellar age at such early epochs. The slope of the relation depends on several dust properties, including production and destruction by supernovae, grain properties, geometry, and the fraction of dust removed by outflows. \citet{langeroodi_24b} find a time delay of about 30 Myr, consistent with SN dust production. We probe a lower mass regime than either work, but still require an inverse relation between \Av\ and sSFR as seen in the bottom left panel of Figure \ref{fig:Av}, pointing to a gradual buildup of dust. We refrain from reporting the slope of the relation due to the limit in sSFR values which particularly impacts our low mass sample.  

\section{Summary and Conclusion} \label{sec:Conclusion}

We have presented a sample of $z>7.5$ galaxies from the JWST Technicolor and CANUCS surveys in order to investigate the amount of dust obscuration in the very early universe. We summarize our main findings as follows:

\begin{itemize}
    \item Including color-selected ELGs by utilizing medium-bands improves the completeness of high-z galaxy samples by reaching fainter UV magnitudes. In our sample of 146 $z>7.5$ galaxies, 15 (10\%) are only selected via the emission line selection, partly due to intrinsically fainter UV and partly due to dust.

    \item Inclusion of medium-bands in SED fitting of high-z galaxies significantly improves accuracy of photometric redshifts and reduces systematic bias in stellar mass and SFR estimates by $\sim 0.2$ and $\sim 0.1$ dex respectively.
    
    \item We observe trends of increasing dust obscuration derived from SED fitting with increasing stellar mass, SFR, and UV-slope, $\beta$, though with significant scatter. 

    \item We use our derived attenuation values to infer the fraction of star formation obscured by dust as a function of redshift. We find that $\sim 50\%$ of all star formation up to $z=9$ is obscured by dust, with the increased obscuration in more massive, heavily star-forming galaxies playing a significant role. The obscuration fraction drops in our highest redshift bin, perhaps pointing to a limit in the timescale in which significant dust can accumulate.
\end{itemize}

In order to verify the findings presented here, we require direct constraints on dust emission or attenuation subject to fewer modeling assumptions. Large samples of typical galaxies at $z>7$ will be time-consuming and difficult to observe with ALMA, and very deep spectra are required to measure Balmer line ratios for galaxies as faint as those in this sample (F277W $\sim$ 30 mag). A possible alternative is to better constrain the rest-frame optical-NIR SED with deep MIRI imaging, which may be more efficient. Additionally, the small volume probed by Technicolor does not allow us to sufficiently sample the high-mass ($M_*>10^{9.5} M_\odot$) galaxy population and precludes more direct comparison with existing ALMA observations. A large-area medium-band JWST survey could make great progress in this regard.

\begin{acknowledgments}
NM, AH, MB, GR, RT, and VM acknowledge support from the ERC Grant FIRSTLIGHT and Slovenian national research agency ARRS through grants N1-0238 and P1-0188. MB  acknowledges support from the program HST-GO-16667, provided through a grant from the STScI under NASA contract NAS5-26555. Based on observations with the NASA/ESA/CSA James Webb Space Telescope obtained from the Data Archive at the Space Telescope Science Institute, which is operated by the Association of Universities for Research in Astronomy, Incorporated, under NASA contract NAS5-03127. DM acknowledges funding from JWST-GO-03362, provided through a grant from the STScI under NASA contract NAS5-03127. This research was supported by grant 18JWST-GTO1 and 23JWGO2A13 from the Canadian Space Agency (CSA), and funding from the Natural Sciences and Engineering Research Council of Canada (NSERC). This research used the Canadian Advanced Network For Astronomy Research (CANFAR) operated in partnership by the Canadian Astronomy Data Centre and The Digital Research Alliance of Canada with support from the National Research Council of Canada the Canadian Space Agency, CANARIE and the Canadian Foundation for Innovation. The Cosmic Dawn Center (DAWN) is funded by the Danish National Research Foundation under grant No. 140. The MAST DOI for CANUCS is \href{https://doi.org/10.17909/ph4n-6n76}{doi:10.17909/ph4n-6n76}.

\end{acknowledgments}

\vspace{5mm}
\facilities{HST(ACS), JWST}

\software{astropy \citep{astropy13},  
          BAGPIPES \citep{carnall18}
          }

\appendix
\section{Systematics from SED Fitting Assumptions}
\begin{figure*}[h!]
\plottwo{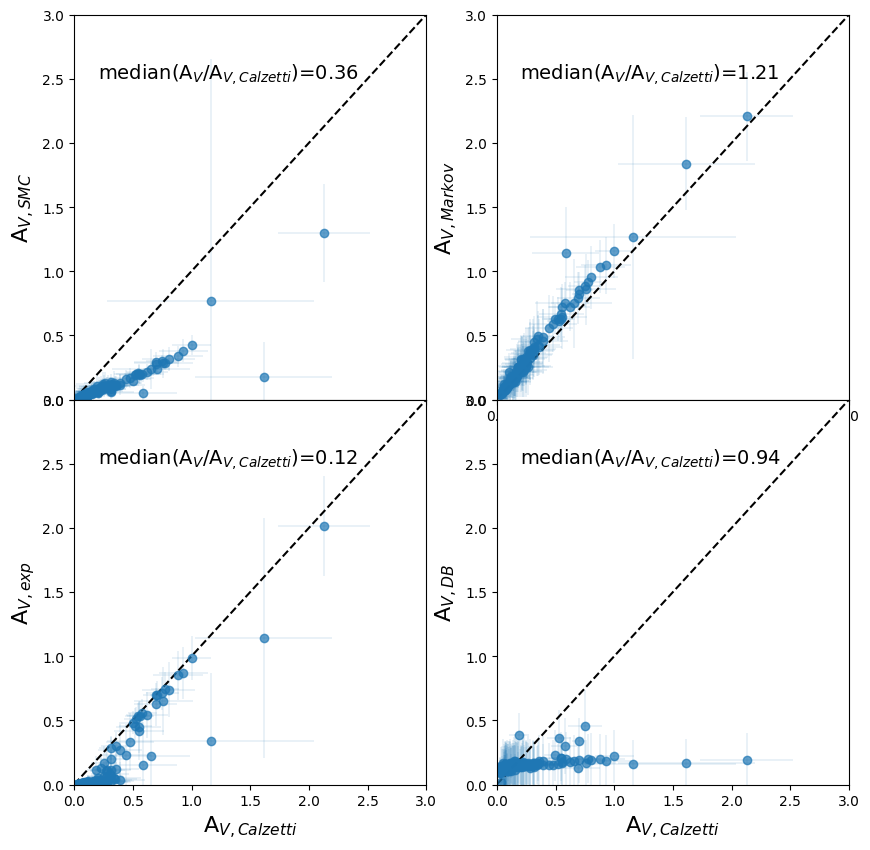}{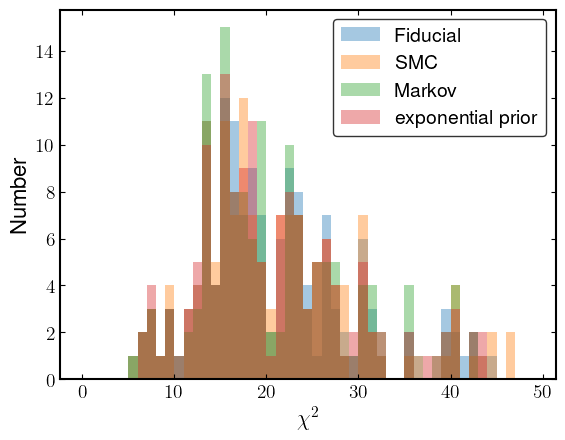}
\caption{Left: Comparison of measured A$_V$ when adopting an SMC \citep{gordon03} attenuation curve (top left) and average attenuation curve of $z=7-10$ galaxies from \citet{markov24} (top right) to that from the \citet{calzetti00} attenuation curve used in this paper. The bottom left shows results when using an exponential rather than flat prior on dust. The bottom right shows the \Av\ derived from SED fitting with the \texttt{DENSEBASIS} code. The median ratio of A$_V$ values for each is indicated. Right: Distribution of $\chi^2$ values for the SED fits for different modeling assumptions as labeled.
\label{fig:Av_comparison}}
\end{figure*}

Recent work by \citet{markov24} has directly measured the median attenuation curve for galaxies at high-z and finds results similar to the Calzetti curve in the $6.3 < z < 11.5$ range, rather than the steeper Small Magellanic Cloud (SMC) curve \citep{gordon03} proposed by others. Here we compare to the median curve for the $7<z<10$ subsample derived using the same methodology to more closely match the sample in this paper. To facilitate comparison with other work, we report our main findings assuming a \citet{calzetti00} curve, but here show the inferred attenuation values for the SMC and empirical \citet{markov24} curves in the left sub-figure of Figure \ref{fig:Av_comparison}. The comparisons expectedly show systematic bias in \Av\ ($\sim 65\%$ smaller for SMC, $\sim 20\%$ larger for the empirical curve). However, the comparisons also show small scatter, suggesting the primary uncertainty is systematic due to the attenuation curve choice and that the dense sampling of the rest-frame UV-optical SED can constrain the amount of dust attenuation given that choice. The bottom left panel shows the results assuming an exponential rather than flat prior on dust. For galaxies with less than $A_V \lesssim 0.3$, many of the fits move close to zero, showing that the models can not discriminate between very little and no dust. The attenuation derived for the most heavily obscured galaxies, however, remains consistent, albeit with larger errors. This suggests the result of the presence of at least some heavily obscured early galaxies is strongly preferred. Finally, the bottom right panel shows the attenuation derived from fitting the sample with the \texttt{DENSEBASIS} SED modeling code, which uses a non-parametric star formation history and here assumes an exponential prior on {\Av}. Almost all the derived attenuation values all fall within a narrow range around $A_V \sim 0.2$, suggesting either minimal impact from dust at these redshifts or that the \Av\ is prior-driven in \texttt{DENSEBASIS} in this case. As an additional check, the right sub-figure of Figure \ref{fig:Av_comparison} shows the $\chi^2$ distributions for each of the \texttt{BAGPIPES} modeling cases are indistinguishable, meaning that none provides a clearly better fit to the data. Discriminating between the assumptions between the fiducial case and those shown here will be difficult and resource-intensive, requiring deep spectroscopic campaigns careful designed to avoid biased populations, or ultra-deep MIRI imaging to constrain the rest-optical emission and reduce the degeneracies between dust attenuation, stellar age, and metallicity.

\bibliography{refs}{}
\bibliographystyle{aasjournal}

\end{document}